\documentclass[lnbip]{svmultln}
\usepackage{makeidx}  % allows for indexgeneration
\usepackage[pdftex]{graphicx}
\usepackage{mdframed}

\begin{document}
\mainmatter              % start of the contribution
\title{Improving Communication\\ in Scrum Teams}
\titlerunning{Improving Communication in Scrum Teams}  % abbreviated title (for running head)
\author{Marvin Wyrich\inst{1} \and Ivan Bogicevic\inst{2} \and Stefan Wagner\inst{2}}
\authorrunning{Marvin Wyrich et al.}   % abbreviated author list (for running head)
\tocauthor{Marvin Wyrich, Ivan Bogicevic, Stefan Wagner}
\institute{AEB GmbH, Stuttgart, Germany\\
\email{marvin.wyrich@aeb.com}
\and
Institute of Software Technology, University of Stuttgart, Germany\\
\email{(ivan.bogicevic|stefan.wagner)@informatik.uni-stuttgart.de}
}
\maketitle              % typeset the title of the contribution
\index{Wyrich, Marvin}  % entries for the author index
\index{Bogicevic, Ivan} % of the whole volume
\index{Wagner, Stefan}

\begin{abstract}        % give a summary of your paper

Communication in teams is an important but difficult issue. In a Scrum development process, we use the Daily Scrum meetings to inform others about important problems, news and events in the project. When persons are absent due to holiday, illness or travel, they miss relevant information because there is no document that protocols the content of these meetings. We present a concept and a Twitter-like tool that improves communication in a Scrum development process. We take advantage out of the observation that many people do not like to create documentation but they do like to share what they did. We used the tool in industrial practice and observed an improvement in communication.

% please supply keywords within your abstract
\end{abstract}
\section{Introduction}
Communication is an essential part of the work in any Scrum team. Many of the everyday events and activities are communicated orally in the Daily Scrum. The larger a team gets the more difficult is the communication.
%However, we conducted two studies which show that these stand-up meetings alone do not give enough support to an individual team member in staying informed about the whole team's activities.
%A company-wide survey was carried out among members of Scrum teams, in which we asked for the ways information is shared and the circumstances in which short stand-up meetings take place.
%In addition we conducted a two-week study in a single Scrum team, letting each individual member track its daily activities.
%This gave us insights in the types of activities and events that take place, how they can be prioritized and which of them get communicated in the Daily Scrum.

%To overcome the recognized deficit in efficient and independent information acquisition, especially after being absent for a while, we developed a tool called \textit{happening}.
%It allows individual activity tracking with the possibility to automatically generate a summarized representation of daily activities and events of the whole team for a selected time period.
%The evaluation of the tool in a multi-country Scrum team approved its usefulness.

%
\subsection{Problem Statement}
Every day, a lot of activities and events happen in agile teams and some of them are really important for other team members to know.
Especially whenever decisions were taken, new tasks emerged or unexpected incidents happened, and team members are not adequately informed it can become problematic.
The Daily Scrum allows members of a Scrum team to keep up with the latest activities of their colleagues.
The problem is that Daily Scrum meetings do not get documented and not every team member takes part in these meetings regularly. We also observed that not every important activity or event is communicated in the Daily Scrum, e.g. due to the fact that there are several other communication channels or most of the team members took part in the event.
So even a documentation of the Daily Scrum would not cover the whole team's activities.
The situation becomes even worse if a team member is absent for a while.
%Within our two-week activity tracking study, there were about ten different and particularly important events that every team member should be aware of.
The consequence is that an individual has to seek for these information in different places and has to ask his or her colleagues.
This costs valuable time of several team members and it is not guaranteed that the information seeker gets informed about every important event.

\subsection{Research Objectives and Contributions}
The objective of our research is to lower the effort of getting a complete overview of what recently happened within a Scrum team.
In particular we want to develop a tool for generating a summary based on activities and events tracked by individual team members.
This summary should contain all relevant information for an individual team member to eliminate the need for most additional information sources.
The tool aims to ensure better communication in large Scrum teams even when developers are absent for a while.

%
%\subsection{Contributions}
%The contribution of our research is a lightweight tool called \textit{happening} for tracking and summarizing internal activities and events in a Scrum team.
%The idea behind \textit{happening} is to improve communication especially in large teams where not every team member regularly takes part in the Daily Scrum.
%Furthermore, it can be used during the Daily Scrum as reminder for participants and as communication medium for absent team members.

\section{Related Work}
One of the most important challenges in large Scrum teams is inter-team coordination \cite{din:moe}. Coordination requires communication both between teams and within a team. \cite{rks} shows in a case study that handling with knowledge over a longer period of time can only be managed with extensive communication. But important events that occur during development could be relevant to others and are often not sufficiently documented or communicated \cite{vis:coo}. Software developers do not update relevant documents, do not see their benefits and wish more automatic generation of documented content \cite{for:let} \cite{lsf}. An example of such a generation is an automatic summarizer for daily scrum meetings proposed by \cite{par}. Nevertheless there is a gap in the current state of the art on how developers of agile teams stay informed when being absent. Our proposed tool tries to fill this gap.

\begin{figure}
	\centering
	\includegraphics[width=\columnwidth,trim={0 0 0 0},clip]{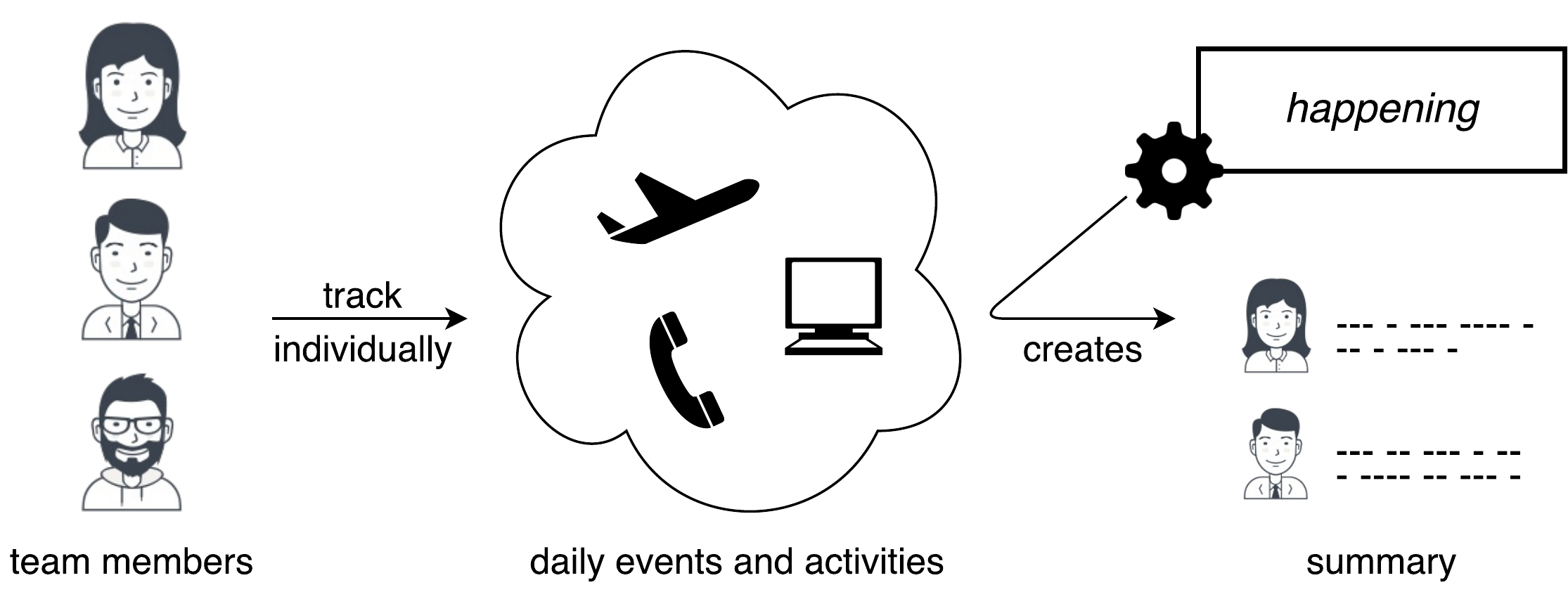}
	\caption[Concept behind happening]{\textit{happening} allows individual members of a Scrum team to efficiently track their activities. The tool then provides a summary of the whole team's activities.}
	\label{fig:happening_concept}
\end{figure}

\section{Concept and Solution}
We described the lack of documentation of daily activities and events and how this negatively influences the communication in Scrum teams.
People usually do not like to create documentation but they do like to share what they did.
As a consequence \textit{happening} serves as a documentation tool that gives an individual team member a way to share his or her experiences in short form, just like they would do on Twitter.
Then \textit{happening} creates the documentation on demand by generating a summarized representation of the  entries, as shown in Fig. \ref{fig:happening_concept}.

The solution consists of two parts: a simple form for inserting individual events and a  
page for viewing the summary for a selected time period.
The latter is shown in Fig. \ref{fig:happening_summary}.
Any event entry consists of a description, a manually selected priority and the date on which the event took place.
The priority is on a scale from one to three and indicates for what period of time the event will be relevant to others.
Thus the priority of an individual event has significant influence on what is shown in the summary if the user wants to hide events that are no longer relevant.
Currently the solution is a stand-alone tool with a web interface and thus can be accessed by its users via any web browser. We also offer an online demo installation where the tool can be tried out without installation.\footnote{https://github.com/MarvinWyrich/happening}

\definecolor{graybordercolor}{RGB}{210,210,210}
\begin{figure}
  \centering
  \begin{mdframed}[linecolor=graybordercolor, linewidth=1px]
  \includegraphics[width=\columnwidth,trim={0 0 0 0},clip]{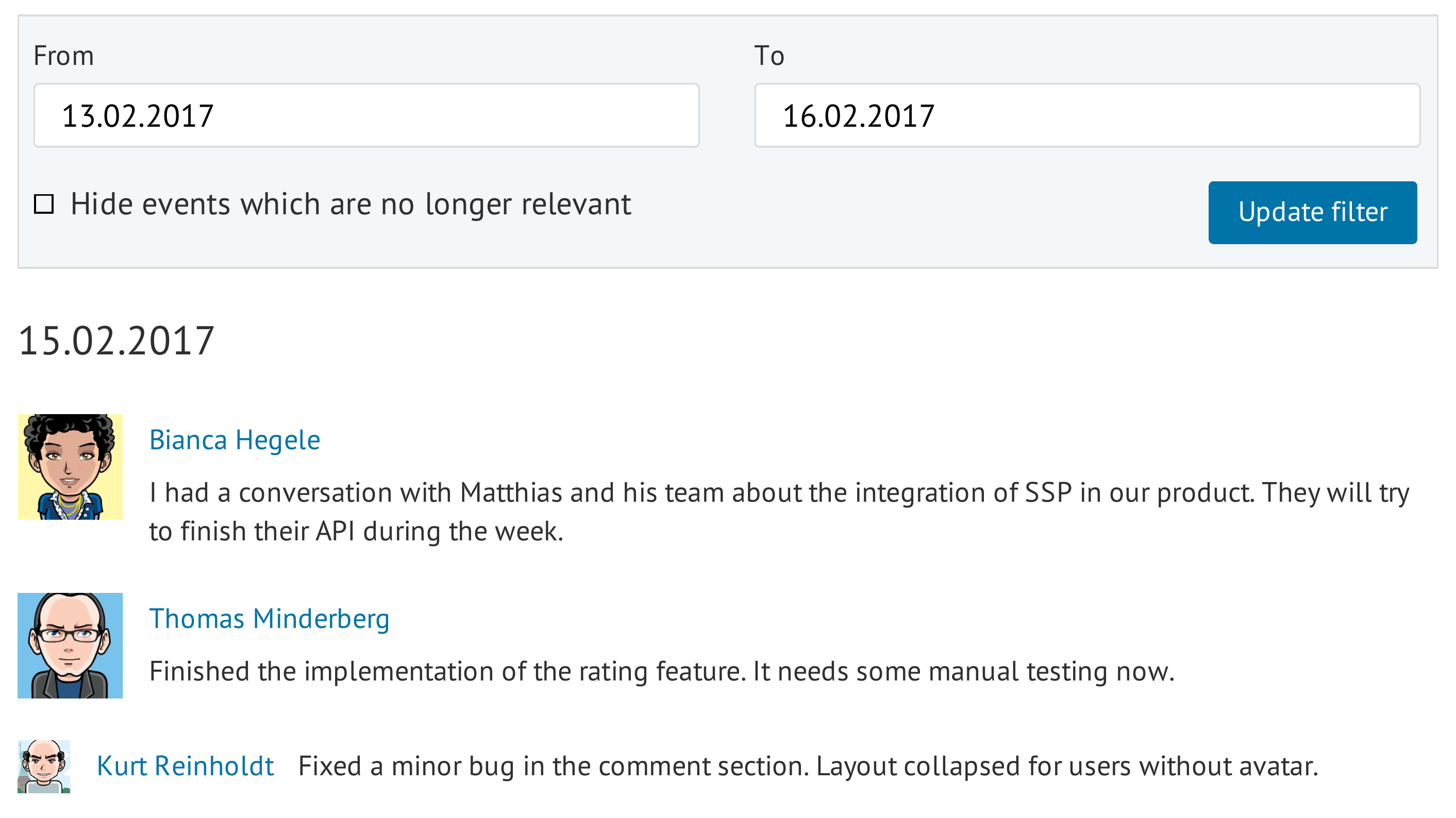}
  \end{mdframed}
  \caption[Summary of events]{Summary of activities and events of a Scrum team for a selected time period. The entry of Kurt Reinholdt was given a lower priority and thus has a smaller avatar.}
  \label{fig:happening_summary}
\end{figure}

\section{Evaluation and Conclusion}
We evaluated the tool in a productive environment of a Scrum team, which used the previously described tool in its day-to-day work.
The team was made up of eight persons, from which one worked from home, one was located in Sweden and the others were at the same office in Germany during the evaluation period.
The team members were told to track their activities and events on a daily basis and to use the summary of the tool in their Daily Scrum.
At the end of the evaluation period the team members gave their anonymous feedback in a questionnaire.

We found that the summarizing presentation of the team's activities is not that useful in the Daily Scrum.
The selected Scrum team was already used to have the JIRA task list opened during the Daily Scrum. So the developers wished to integrate happening as a plugin in JIRA to not have two tools open at the same time.
However, seven out of eight participants said that \textit{happening} was useful outside of the Daily Scrum and they think it would be a great help after a longer period of absence.
Moreover, every participant responded that the use of \textit{happening} would be worthwhile to his or her team, preferably integrated in existing Scrum tools.

The concept and tool worked well in practice and helped improving the communication in agile teams. The costs for using the tool are low as sharing the important information goes fast.

\vspace{-0.2cm}
\section*{Acknowledgements}
\vspace{-0.1cm}
We want to thank AEB GmbH who made this work possible and who provided the persons for the tool evaluation. We also thank Fujitsu Next who supported this work with a 5.000 Euro grant and the first price of its \textit{Agile IT}-Award 2016.


%
\end{document}